\documentclass{article}
\usepackage[dvips]{graphicx}
\begin{document}
\title{Tip Oscillation of Dendritic Patterns in a Phase Field Model}
\author{Hidetsugu   Sakaguchi and Seiji  Tokunaga\\
Department of Applied Science for Electronics and Materials,\\ Interdisciplinary Graduate School of Engineering Sciences,\\
Kyushu University, Kasuga, Fukuoka 816-8580, Japan}
\maketitle
\begin{abstract}
We study dendritic growth numerically with a phase field model.
Tip oscillation and regular side-branching are observed in a 
parameter region where the anisotropies of the surface tension and the 
kinetic effect compete.  The transition from a needle pattern to a dendritic pattern is conjectured to be a supercritical Hopf bifurcation.
\end{abstract}
\section{Introduction}
Crystal growth  has been intensively studied 
as a problem of pattern formation far from equilibrium.\cite{rf:1,rf:2} Dendrites are typical growth patterns found in 
crystal growth exhibited by most metal alloys and some plastic crystals.\cite{rf:3,rf:4} The phase field model, which has the form of the Ginzburg-Landau equation coupled 
with a diffusion equation, is a useful model for such growth patterns.\cite{rf:5,rf:6,rf:7,rf:8}
For melt growth, the Ginzburg-Landau equation represents the 
dynamics of a phase transition from a liquid phase to a solid phase, and 
it is coupled with a heat conduction equation for the latent heat generated at the growing interface.

The standard theory of crystal growth predicts that a needle 
crystal selected by the solvability condition is stable, and 
side branches are produced owing to thermal noise.\cite{rf:9,rf:10}
However, there are some experiments suggesting that side branches 
may appear as a result of regular tip oscillations.\cite{rf:11,rf:12,rf:13,rf:14}
Liu and Goldenfeld investigated the linear stability of a steady needle crystal using the boundary layer model and found an instability in a certain range of  parameter values of the surface tension and kinetic coefficients.\cite{rf:15}
Ihle numerically investigated the competition between the kinetic and surface tension anisotropy in dendritic growth using a fully dynamical front-tracking method and found oscillatory growth.\cite{rf:16} 
However, the numerical results do not seem to exhibit clear oscillation, probably owing to numerical noise resulting from the discretization. 
Such numerical noise effects seem to be small in the phase field model. 
We have numerically investigated the competition between the kinetic and surface tension anisotropy using a phase field model and found that the growth direction changes from $\langle 10\rangle$ to $\langle 11\rangle$ as the kinetic coefficient is changed.\cite{rf:17}  This is a transition from a surface tension dendrite to a kinetic dendrite.  Such transitions have been observed in experiments of viscous fingering and NH$_4$Cl crystal.\cite{rf:18,rf:19}      
In a previous work, we found that complicated patterns grow from small round seeds in an intermediate parameter region where the two anisotropies compete. However, we could not find clear limit cycle oscillation of the tip growth. In this work, we consider the situation in which the anisotropy parameter is changed slowly, in a stepwise manner, and find  a transition from a steady needle pattern to a tip-oscillating dendrite. 

\section{Tip oscillation of dendritic patterns in the phase field model}
The phase field model in two dimensions is written 
\begin{eqnarray}
\tau(\theta)\partial_t p&=&\{p-\lambda u(1-p^2)\}(1-p^2)\nonumber\\
& &+\partial_x\{W(\theta)^2\partial_x p-W(\theta)W'(\theta)\partial_y p\}\nonumber\\
& &+\partial_y\{W(\theta)^2\partial_y p+W(\theta)W'(\theta)\partial_x p\},\nonumber\\
\partial_tu&=&D\nabla^2u+\partial_t p/2,
\end{eqnarray}
where $p(x,y,t)$ is the order parameter, $\lambda$ is a coupling constant, and $\tau(\theta)$ is the time constant for the order parameter, with 
$\theta=\arctan(\partial_yp/\partial_x p)$ the angle between the 
normal direction of the contour of constant $p$ and the $x$ axis. 
The solid and liquid phase correspond, respectively, to $p=1$ and $p=-1$. 
The normalized temperature is denoted by $u(x,y,t)=(T-T_M)/(L/C_p)$ where $T(x,y,t),T_M,L$ and $C_p$ are, respectively, the temperature, the melting temperature, the latent heat and the specific heat.  
The diffusion constant for $u$ is denoted by $D$, and  $W^2(\theta)$ is an anisotropic  diffusion constant for the 
order parameter.
The generalized Gibbs-Thomson condition 
(the Wilson-Frenkel formula), 
\begin{equation}
u_i=-d_0(\theta)\kappa-\beta(\theta)v_n,\end{equation}
is assumed for the sharp interface model, where $u_i$ is the normalized temperature at the interface, $d_0(\theta),\kappa,\beta(\theta)$ and $v_n$ denote, respectively, the anisotropic capillary length, the interface curvature, the anisotropic kinetic coefficient and the normal interface velocity. 
Karma and Rappel derived the values of $d_0(\theta)$ and $\beta(\theta)$ 
in the sharp-interface limit to the phase field model as \cite{rf:8}
\begin{eqnarray}
d_0(\theta)&=&\frac{I}{\lambda J}\{W(\theta)+W"(\theta)\},\nonumber\\
\beta(\theta)&=&\frac{I}{\lambda J}\frac{\tau(\theta)}{W(\theta)}\left [1-\lambda\frac{W^2(\theta)}{2D\tau(\theta)}\frac{K+JF}{I}\right ],
\end{eqnarray}
where $I=2\sqrt{2}/3,J=16/15,K=0.13604$ and $F=\sqrt{2}\ln 2$.
If $\tau(\theta)=W^2(\theta)$ and the coupling constant $\lambda$ is assumed to be $\lambda=(2ID)/(K+JF)$, the anisotropic kinetic coefficient becomes zero, and the standard 
Gibbs-Thomson condition is realized.
Karma and Rappel performed numerical simulations for this case, and obtained dendrites with tip velocities and tip shapes that agree well with the theory based on the solvability condition.
The growth law $v\rho^2=$const., where $v$ is the steady growth velocity of the tip and $\rho$ is the tip radius, is then satisfied.

We performed numerical simulations of the phase field model 
with an anisotropic kinetic coefficient. We found that if the kinetic effect is included, the tip radius does not become sharper and sharper,  
in contrast to the case in which there is no kinetic effect and the relation $v\rho^2$=const. is not satisfied, when the supercooling $\Delta$ is close to 1.\cite{rf:20}  
In this paper, we study the competition of the surface tension 
anisotropy and the kinetic coefficient anisotropy. 
Four-fold rotational symmetry is assumed for the anisotropy: 
$W(\theta)=1+e_4\cos(4\theta)$ and $\tau(\theta)=W(\theta)\{1+e_k\cos(4\theta)\}$, 
where $e_4$ and $e_k$ denote the anisotropy parameters of the surface tension and the 
kinetic coefficient. 
The coupling constant and the diffusion constant are chosen as $\lambda=1.9ID/(K+JF)$ and $D=2$. This choice of parameter values corresponds to the anisotropy coefficients $d_0(\theta)\propto 1-15e_4\cos(4\theta)$ and $\beta(\theta)\propto 0.05+(e_k-0.95e_4)\cos(4\theta)$ in the sharp interface limit, owing to Eq.~(3).\begin{figure}[htb]
\begin{center}
\includegraphics[width=8cm]{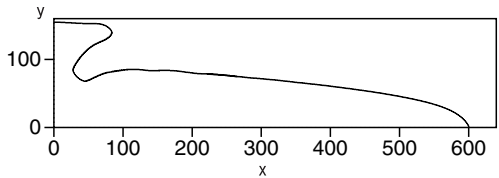}
\caption{Needle pattern exhibited by the phase field model with  four-fold 
rotational symmetry for $e_4=0.06, \Delta=0.7$ and $e_k=0.07$.
} 
\label{fig:1} 
\end{center}
\end{figure} 
\begin{figure}[htb]
\begin{center}
\includegraphics[width=8cm]{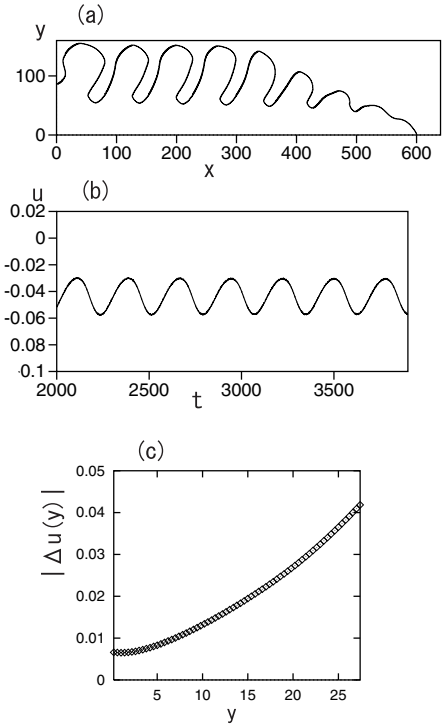}
\caption{(a) Dendritic pattern with regular side branches at $e_4=0.06,\Delta=0.7$ and $e_k=0.08$. 
(b) Time evolution of the interface temperature $u(y,t)$ for $y=20$. 
(c) Amplitude $|\Delta u(y)|$ of the oscillation as a function of $y$.
} 
\label{fig:2} 
\end{center}
\end{figure} 
\begin{figure}[htb]
\begin{center}
\includegraphics[width=6cm]{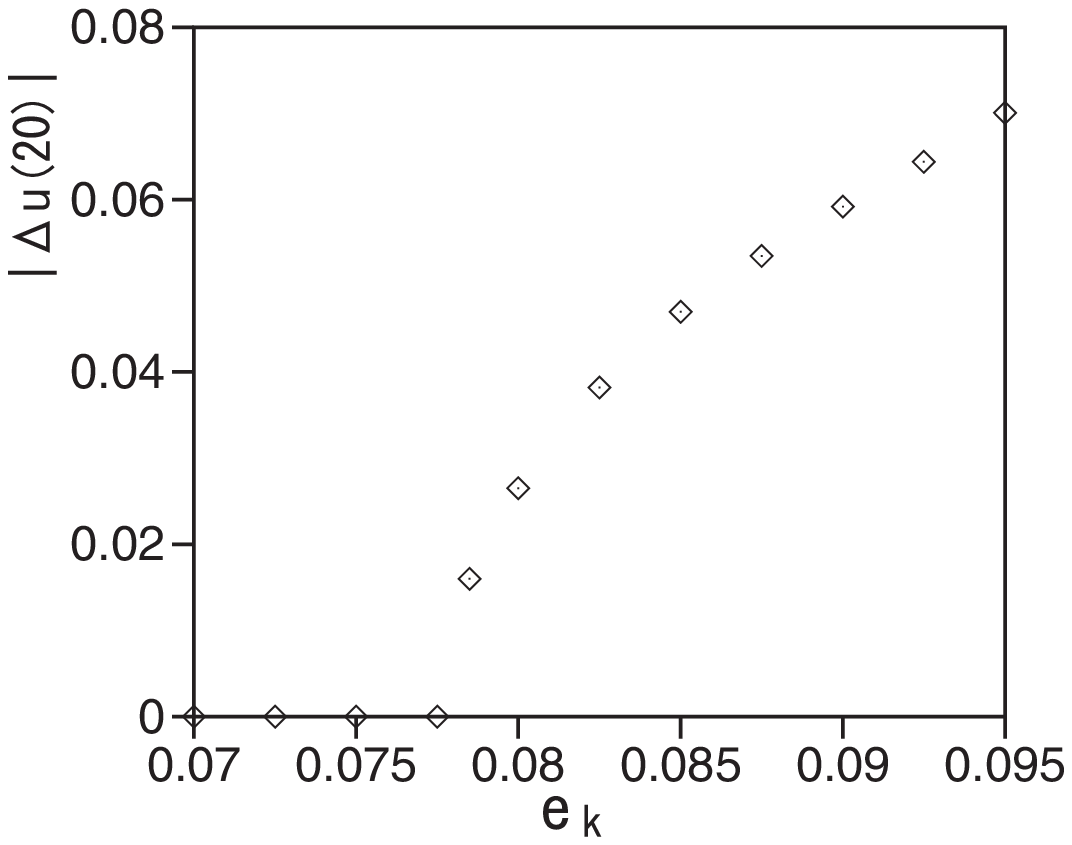}
\caption{Amplitude of the oscillation $|\Delta u(y)|$ at $y=20$ 
as a function of $e_k$.
} 
\label{fig:3} 
\end{center}
\end{figure} 
 
When $e_k$ is larger than 0.95$e_4$, the $\langle 11\rangle$ growth becomes favorable,  
owing to the kinetic effect.   
Numerical simulations were performed using the finite 
difference method with $\Delta x=0.4$ and $\Delta t=0.015$. 
The system size was $L_x\times L_y=640\times160$.
The uniform undercooling $u=-\Delta$ was employed as the initial condition. 
Mirror-symmetric boundary conditions  were used at $x=0$ and $y=0,L_y$ and the fixed boundary conditions $p=-1$ and $u=-\Delta$ were used at $x=L_x$.
The parameters $\Delta$ and $e_4$ were fixed as $\Delta=0.7$ and $e_4=0.06$, and the parameter $e_k$ for the kinetic anisotropy was changed step by step.

The process used in our numerical simulation is as follows.
If the tip of the dendritic pattern reaches $x=600$, the growth for the parameter $e_k$ is stopped, and the numerical data for the order parameter and the temperature are saved.  
The order parameter and the temperature profiles in the tip region ($\tilde{p}(x,y),\,\tilde{u}(x,y)$ for $x>520$) are used as the initial conditions for the parameter $e_k$ at the next step, that is, $p(x,y,0)=\tilde{p}(x-520,y),\,u(x,y,0)=\tilde u(x-520,y)$. 
\begin{figure}[htb]
\begin{center}
\includegraphics[width=8.5cm]{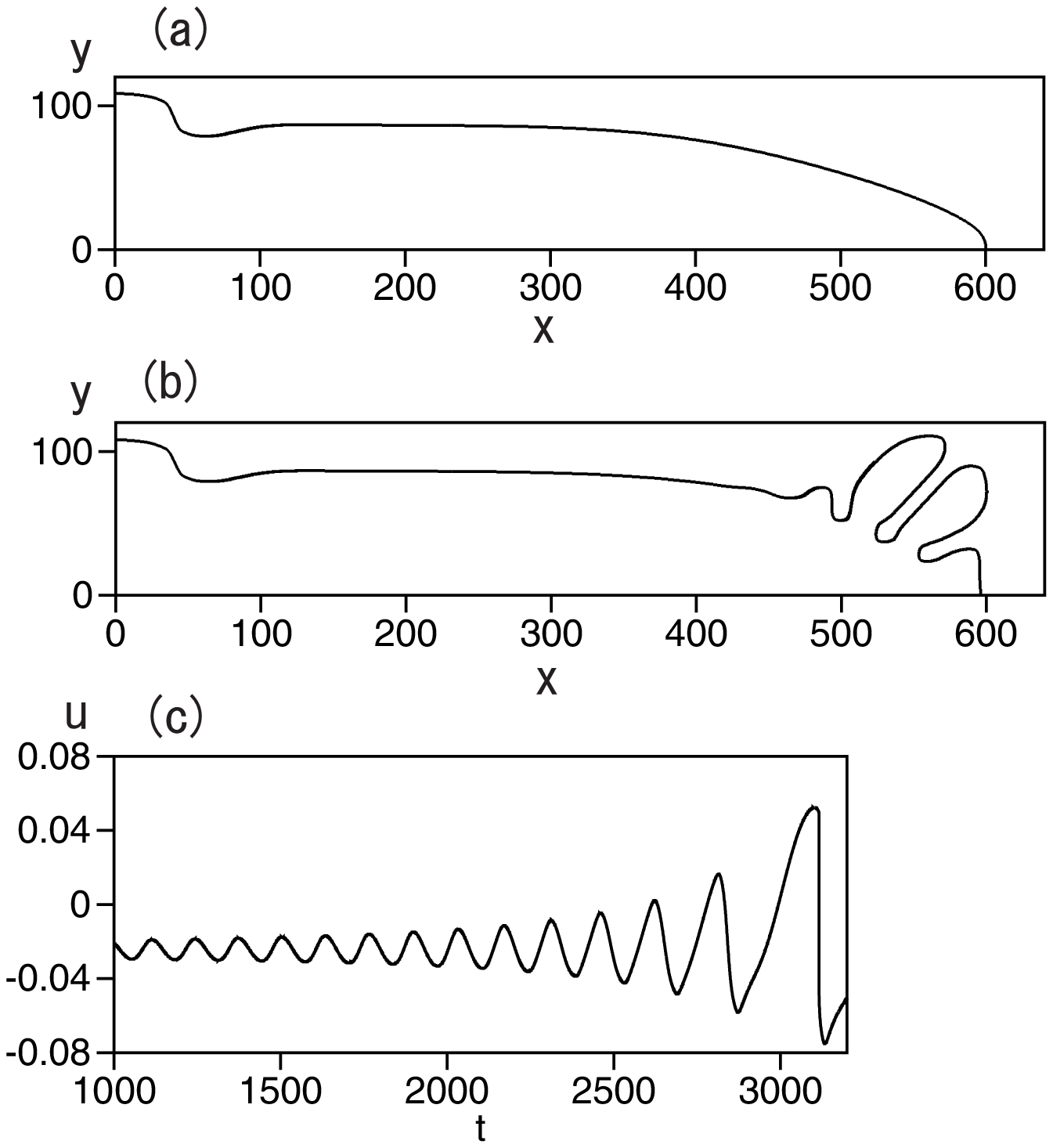}
\caption{(a) Needle pattern for $e_4=-0.06, \Delta=0.7$ and $e_k=-0.088$.  
(b) Snapshot of the pattern for $e_4=-0.06, \Delta=0.7$ and $e_k=-0.085$.
(c) Time evolution of the interface temperature $u(y,t)$ at $y=20$.} 
\label{fig:4} 
\end{center}
\end{figure} 
\begin{figure}[htb]
\begin{center}
\includegraphics[width=8cm]{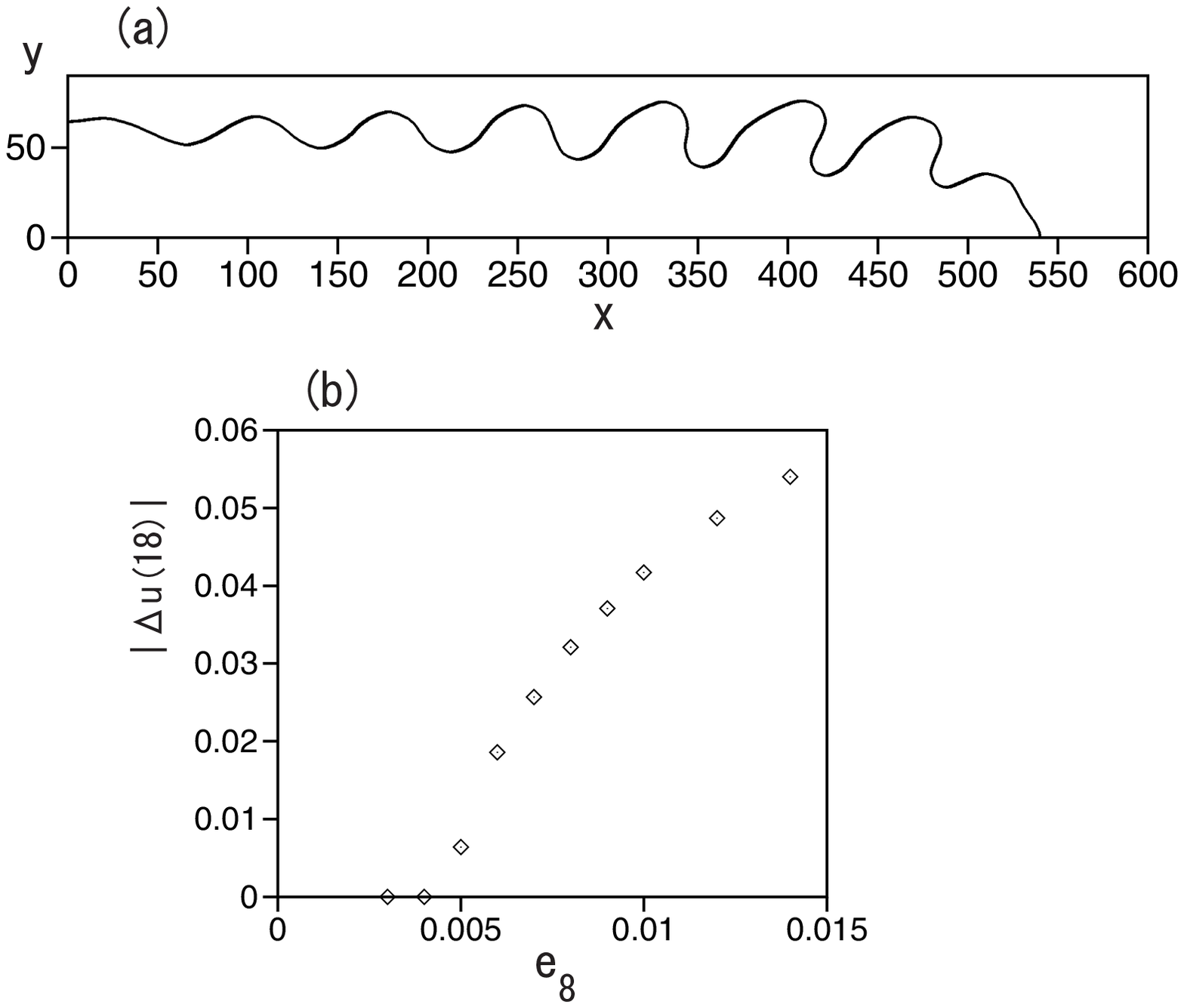}
\caption{(a) Dendritic pattern for the phase field 
model with eight-fold symmetry for $e_8=0.01$ and $\Delta=0.65$. 
(b) Amplitude $|\Delta u(y)|$ of the oscillation of the interface temperature  at $y=18$ as a function of $e_8$.
} 
\label{fig:5} 
\end{center}
\end{figure} 
Figure 1 displays a snapshot of a needle pattern for $e_k=0.07$.  
This needle pattern is growing in the $\langle 10\rangle$ direction, owing to the surface tension anisotropy. 
As the parameter $e_k$ is increased, the growth direction changes from the $\langle 10\rangle$ direction to the $\langle 11\rangle$ direction due to the kinetic effect. 
In region of  intermediate values of the parameter $e_k$, a dendritic pattern with side branches appears. 
Figure 2(a) displays a dendritic pattern for $e_k=0.08$.
Regular side branching is seen. The interface temperature $u(y,t)$ is evaluated for each $y$ as the temperature at the position $(x,y)$ where $p(x,y)=0$ is satisfied. For this parameter value, the average tip velocity is approximately $v=0.265$, and the diffusion length $l=2D/v\sim 15$.  The system size is sufficiently larger than the diffusion length.  Figure 2(a) displays a snapshot pattern at the instant that the tip reaches the position $x=600$, and the effect of the boundary at $x=L_x=640$ is not yet important.  
Figure 2(b) displays the interface temperature at $y=20$. 
The interface temperature and the curvature oscillate regularly  at this value of the parameter.  Figure 2(c) displays the amplitude of the oscillation $|\Delta u(y)|=\max u(y,t)-\min u(y,t)$ as a function of $y$, where $\max u(y,t)$ and $\min u(y,t)$ are, respectively,  the maximum and minimum values in the oscillatory time evolution of $u(y,t)$. This amplitude increases monotonically, but, it does not seem to fit a curve such as $\exp(cs^{1/4})$ or $\exp(cs)$, where $s$ is the arc length along the interface and $c$ is a constant, as is expected from some theories.
It seems, rather, to fit a linear curve, $cs$, although we do not understand 
the reason for this. 
Figure 3 displays the amplitude of the oscillation $|\Delta u(y)|$ at $y=20$ 
as a function of $e_k$.  As seen, there is a transition 
from a stationary state to oscillatory growth, and the amplitude increases continuously from zero; that is, there is no jump at the transition.  
The transition is therefore concluded to be a supercritical Hopf bifurcation. 
This limit cycle process can be interpreted as follows. The surface tension dendrite grows in the $\langle 10\rangle$ direction. The side branches grow in the $\langle 11\rangle$ direction, owing to the kinetic effect. Then, the dendritic tip becomes round, the radius of curvature of the tip region increases, and the growth velocity in the $\langle 10\rangle$ direction decreases. As the velocity is decreased, the kinetic effect, which is proportional to the growth velocity,  becomes weak, and the surface tension effect become relatively large. Then, growth in the $\langle 10\rangle$  becomes dominant, and the velocity in the $\langle 10\rangle$ direction increases. As the growth velocity  increases, the kinetic effect becomes relatively large, and the side branches grow in the $\langle 11\rangle $ direction again.  In this way, a negative feedback acts between the tip velocity in the $\langle 10\rangle$ direction and the kinetic effect,  and a stable limit cycle appears.

Next, we consider the transition for kinetic dendrites.
If $e_k<0$, $e_4<0$ and $e_k$ is sufficiently smaller than $e_4$, 
a steady needle pattern grows in the $\langle 10\rangle$ direction, owing to the kinetic anisotropy. As $e_k$ remaining less than 0 is increased, the kinetic anisotropy becomes relatively weak, and the surface tension becomes dominant. Then a transition from $\langle 10\rangle$ growth to  $\langle 11\rangle$ growth occurs.  In the simulations, the parameters $\Delta$ and $e_4$ were fixed as $\Delta=0.7$ and $e_4=-0.06$, and the parameter $e_k$ was changed stepwise.  The system size was $L_x\times L_y=640\times 120$. Figure 4(a) displays a needle pattern for $e_k=-0.088$. The needle pattern is marginally stable for $e_k\sim -0.087$, and it becomes unstable for $e_k>-0.086$. 
Figure 4(b) displays the snapshot of a pattern for $e_k=-0.085$.  Side branches develop in the tip region, and the growth of the side branches does not saturate. Figure 4(c) displays the temporal evolution of the interface temperature $u(y,t)$ at $y=20$. The amplitude of the oscillation of the interface temperature grows but does not saturate.   The side branches develop and they become main branches if $L_y$ is sufficiently large. 
The negative feedback does not act in this case. 
If the side branches grow, owing to the surface tension effect, the dendritic tip becomes more round and the growth velocity in the $\langle 10\rangle$ direction decreases. Then the kinetic effect becomes further weakened,  the surface tension effect becomes more dominant, and the growth in the $\langle 11\rangle$ direction  does not saturate.

Elucidation of the oscillatory growth caused by the competition between two types of anisotropies is the main result of this paper. However, we also demonstrate a transition to the tip oscillation of dendritic growth, where the surface tension anisotropy  has  eight-fold rotational symmetry.\cite{rf:17,rf:21}  In a previous work, we showed the existence of tip oscillation in such a system, but have not demonstrated the transition from the needle pattern to  the oscillation.
The parameter dependence of the oscillatory growth is observed when $\tau(\theta)=W^2(\theta)$, where  $W(\theta)=1+e_8\cos8\theta$. The coupling constant $\lambda$ is chosen to be $\lambda=(2ID)/(K+JF)$, so that  
the kinetic effect can be neglected.  The numerical simulation was performed using the finite difference method with $\Delta x=0.3$ and $\Delta t=0.01$. 
The system size was $L_x\times L_y=600\times 90$ and the undercooling was fixed as $\Delta=0.65$. 
Figure 5(a) displays a regularly oscillating dendrite for $e_8=0.01$. 
The interface temperature is seen to exhibit limit cycle oscillation. 
Figure 5(b) displays the amplitude of the oscillation of the interface temperature $|\Delta u(y)|$ at $y=18$ as a function of $e_8$. 
A supercritical bifurcation occurs at $e_8\sim 0.045$ for this value of the undercooling. 
There are few real crystals that possess eight-fold rotational symmetry.  However, it is not so unusual for the surface tension to be expressed as 
$d(\theta)=1+e_4\cos4\theta+e_8\cos8\theta$, in which case, the rotational symmetry is  four-fold,  but the crystal has a component of eight-fold symmetry. 
As the parameter $e_4(>0)$ is increased for a fixed value of $e_8>0$,  $\langle 10\rangle$ growth becomes dominant in comparison with $\langle 11\rangle$ growth. Tip oscillation appears for a small value of $e_4$, but,  
its amplitude decreases as $e_4$ is increased. 
Tip oscillation is not observed for $e_4>0.016$ when $e_8$ is fixed to 0.01. (The transition to non-oscillatory growth seems to be discontinuous, with a jump, although this is not entirely clear.) 
It appears that the competition between $\langle 10\rangle$ growth and $\langle 11\rangle$ growth  may be important in the realization of tip oscillation. 
\section{Summary and discussion}
We have found a deterministic transition from a needle pattern to  a regularly oscillating dendrite in the phase field model. 
This was found in surface tension dendrites with  four-fold rotational symmetry and competitive kinetic anisotropy. 
A similar supercritical transition was found for surface tension dendrites with eight-fold rotational symmetry without the kinetic effect. Regularly oscillating  surface tension dendrites with four-fold or six-fold rotational symmetry without the kinetic effect have not yet been found in this model.   
Regularly oscillating dendrites with side branches possess mirror symmetry as a results of the method used in our numerical simulation.  
The boundary condition is not, however, essential for the oscillation. 
We have confirmed that a mirror-symmetric dendrite grows naturally in a simulation  in which a small needle pattern including a small antisymmetric perturbation is placed in the middle of the left side of the $L_x\times 2L_y$ box as the initial conditions.
These results are not consistent with the linear stability analysis of Liu and Goldenfeld and the simulation of Ihle. They predict oscillatory growth with antisymmetric side branches for both surface tension dendrites and kinetic dendrites.  The detailed mechanism of the oscillatory growth is not yet well understood, and its elucidation is left as a future study.

\section*{Acknowledgements}
We would like to thank  Professors H.~Honjo and S.~Ohta for valuable discussions. 

\end{document}